\documentclass[10pt,twocolumn]{article}
\usepackage{amsmath,amssymb}
\usepackage{graphicx}
\usepackage{cite}
\usepackage{geometry}
\begin{document}
\makeatletter
\twocolumn[
 \begin{@twocolumnfalse}
 \begin{center}
{\Large\bfseries Fresnel's Mechanical Legacy Recovered:\\ The Drag Coefficient from Carried Compliance,\\ and the Two Laws Bubble Acoustics Separates\par}
\vskip 8pt
{Shiva Meucci\par}
\vskip 2pt
{\small\itshape Independent Researcher, Las Vegas, NV, USA\\
E-mail: bmeucci@gmail.com \quad ORCID: 0009-0006-3300-4480\par}
\vskip 4pt
{\small August 20, 2026\par}
\end{center}
\vskip 2pt
\begin{center}\begin{minipage}{0.86\textwidth}
{\small\noindent\textbf{Abstract.}\enspace Fresnel derived the drag coefficient of moving transparent matter from a mechanical argument in 1818, and a century of interferometry confirmed it exactly as he gave it;  We show that the mechanical underpinning he sought was workable in principle and wrong in one word: density, where the mechanics says compliance. The complete first-order coefficient follows from a single constitutive calculation, each component responding along its own path at the frequency it samples:
\[
f(\omega) \;=\; \frac{C_{\mathrm{soft}}}{C_{\mathrm{total}}} \;+\; \frac{\omega}{2}\,\frac{C_{\mathrm{soft}}'(\omega)}{C_{\mathrm{total}}}\,.
\]
The coefficient contains two laws. The first is the law Fresnel wrote: the non-dispersive part, $1-1/n^2$, equals the modified medium's share of the wave-accessible compliance, $\chi/(1+\chi)$: his excess hypothesis made exact. The second is a law of resonance, the term Lorentz derived and Zeeman confirmed. We prove the resonance law is universal mechanics: sound in a bubbly liquid obeys it in the same form and prefactor, eleven orders of magnitude below, invariantly across the analyzed two-fluid closures (Appendix~A). The same analysis shows the share law is the carried-response term: present when the compliant channel stores and delivers in its moving frame, absent from every closure that severs response from carrier. Closure-bound cavities (geometry fixed by phase closure of the medium's waves) provide that custody by construction. One transport structure thus appears three times: light in moving glass carries both terms, the drifting bubbly liquid isolates the resonance term by slipping, and Lorentzian velocity composition reproduces the share. The direction of derivation among the three is established: the mechanics generates what the kinematics reproduces; the composition is itself derived as the cavities' closure condition in a companion paper. A documentary history traces how interpretation fused the two laws, Fresnel to Zeeman.}
\end{minipage}\end{center}
\vskip 10pt
 \end{@twocolumnfalse}]
\makeatother

\section{Introduction}

Fresnel predicted the partial drag of light by moving matter from a mechanical argument; the experiments returned his coefficient exactly as predicted; and the argument's mechanical content, we will show, still holds---all of it. What a century of interpretation lost was not the mechanics but the anatomy: the coefficient contains two laws, fused by the relativistic media in which optics measures them, and a nineteenth-century transport system (sound in bubbly water) is the instrument that separates them and certifies each as mechanical in its own register.

The central result can be stated at once, in plain terms, and it is elementary. Let each component of a medium respond to the pressure it actually experiences, along its own path, at the frequency it actually samples. That single constitutive calculation, classical continuum mechanics with no relativistic input, yields Fresnel's complete first-order coefficient: the non-dispersive share as the carried-compliance fraction, the dispersive term as the Doppler-sampled derivative of the compliant response (Section~2, Appendix~A). This much is established outright and stands on its own. Everything that follows examines the same result from further directions: which physical circumstance separates the  two terms: the bubbly liquid does, observably, the moment its bubbles slip; what the separation proves about each term; and how a century of interpretation lost the mechanical reading.

The first law is the one Fresnel wrote. The non-dispersive part of the coefficient, $1-1/n^2$, is algebraically the modified medium's \emph{share} of the total wave-accessible compliance, $\chi/(1+\chi)$ (Section~2.1): his excess hypothesis made exact, with compliance in the role he assigned to density. The share takes that value when the compliant response is carried  with the flow, stored and delivered in the moving frame; von Laue's 1907 derivation reproduces it from velocity composition \cite{vonlaue1907}, and the two-fluid analysis shows what removes it: sever the response from its carrier (inclusions slipping through a resting liquid) and the residues, of order the void fraction, never take the share form. The share law is thus the carried-compliance fraction, restored the moment the response is carried with the flow (Appendix~A); Lorentzian composition reproduces the same value at first order, and the companion paper (Paper~I of this pair, submitted with this one) derives that composition mechanically, proving that Lorentz--FitzGerald contraction and period dilation are the unique deformation preserving spherical-harmonic phase closure for resonant structures moving through a mechanical wave medium \cite{meucci2026cavity}. Closure means, literally, that the waves circulating inside the cavity return in phase with themselves; the cavity persists only while they do, so a moving cavity must deform or cease to exist. Cavities bound by phase closure of the medium's waves compose velocities Lorentzianly; cavities whose geometry is fixed instead by force balance (a gas bubble, its radius set by gas pressure against surface tension) drift free of that constraint. Far from dissolving Fresnel's mechanics, the kinematic derivation is the part of the mechanics that runs through  measurement itself, and the bubble is the isolating case that shows it.

The second law is the one Lorentz added. The dispersive part of the coefficient equals $(\omega/2)\,C_{\mathrm{soft}}'(\omega)/C_{\mathrm{total}}$: the frequency-derivative of the compliant component's responsiveness, over the total compliance of the medium. This law is universal mechanics, in the sense made precise in Appendix~A: invariant across a two-parameter family of two-fluid closures and across Galilean and Lorentzian kinematics alike.  Light obeys it; it is precisely the term Lorentz derived from moving molecular oscillators in 1895 \cite{lorentz1895} and Zeeman confirmed interferometrically, color by color \cite{zeeman1915}, and we prove in this paper (Section~2, Appendix~A) that sound in a bubbly liquid whose bubbles drift through the resting liquid obeys it in exactly the same form, with exactly the same prefactor, eleven orders of magnitude below in frequency, and independently of the two-fluid closure employed.

This paper is, deliberately, a study in the history of transport phenomena, because the fusion of the two laws has a documented genealogy, and because the historical actors' own vocabulary preserved the distinctions that the later kinematic recasting erased. Fizeau, announcing the confirming experiment, recommended adopting Fresnel's hypothesis ``or at least the law which he found'', severing law from mechanism in the founding report \cite{fizeau1851}. Veltmann stated the chromatic structure of the coefficient in participation language (the rate at which the light-motion \emph{partakes} of the medium's motion), and detached the coefficient from every drag ontology \cite{veltmann1870}. The famous multiple-ethers reductio, we will show, has force only against the density ontology Fresnel's letter happened to fix; under the compliance reading the chromatic structure is not an absurdity but the resonance law itself. Our approach follows the historical method of neoclassical interpretation \cite{meucci2018}: demonstrate the mathematical structure, then trace the conceptual choices that obscured it. The question of what a kinematic derivation shows about underlying dynamics remains actively contested in the foundations literature \cite{bell1976,brown2005,janssen2009}; the two-law anatomy contributes a concrete exhibit in which the mechanically universal part and the kinematically Lorentzian part of one celebrated coefficient are separated by experiment-grade derivation.

A scope note carries over from earlier stages of this program: our claim is mechanical sufficiency, not ontological uniqueness: the paper and its companion supply a generating construction of the law at its order, without asserting that nature admits no other, and with the division of labour made exact. The present paper supplies the transport law and the historical analysis; the companion supplies the closure theorem. Together they address the two halves of the constructive question: how waves are carried by moving structured media, and how wave-organized structures deform when moving.

The paper is organized as follows. Section~2 states the two laws, proves the acoustic separation, and equips the reader with the three distinctions on which everything turns. Section~3 reconstructs the interpretive history, 1818--1915, as the genealogy of the fusion. Sections~4 and~5 develop the physical principle (waves couple to the most compliant component) and its extension to atomic polarization. Sections~6--9 return to the historical record: the Michelson--Morley null, von Laue's theorem, contraction, and the two compensations. Section~10 discusses how the historical conflations reproduce themselves in modern analysis, and how the two-law division prevents them. Section~11 concludes. Appendix~A contains the derivation.

\section{The two laws, stated and separated}

Three transport settings share one first-order template:
\begin{equation}
v_{\mathrm{obs}}(\omega,\theta) = v(\omega) + \left[\,f_{0}(\omega) + f_{\mathrm{res}}(\omega)\,\right] U \cos\theta + O\!\left((U/v)^2\right),
\label{eq:template}
\end{equation}
where $v(\omega)$ is the wave speed in the medium at rest, $U$ the drift velocity, and the coefficient is written from the outset in its two-part anatomy: a non-dispersive part and a resonance part. This section defines both, proves where each does and does not hold, and is the paper's core; everything after it is development and history.

One principle organizes everything in this section: a wave couples to the most compliant component  of its medium: the part with the most give. In bubbly water the compliant component is the gas, visibly; in transparent matter it is the electronic structure, as polarizability; in the constructive model developed later, the vacuum's substrate itself. The two laws below are this single principle in two appearances of one calculation. The method throughout is that of hydrodynamic analog physics \cite{unruh1981,barcelo2011,bush2015}: a laboratory fluid realizes the mathematical structure of the target system, extended here from kinematics and effective metrics to the transport coefficient itself. At rest, the compliant share of the medium's total compliance sets the wave speed: the share identity of Section~2.1. In motion, let each component respond to the pressure it actually experiences, along its own path and at the frequency it actually feels: the carried compliance then contributes its share of the drift (the non-dispersive law of Section~2.3) and its Doppler-sampled response contributes  the derivative term, the resonance law of Section~2.2. One constitutive calculation, both terms, at first order (Appendix~A). The inclusion-drift closures analyzed alongside show when the share disappears: sever the response from its carrier and the share goes with it, leaving the resonance  term alone: the separation the bubbly liquid makes observable. Compliance is the mechanism throughout; the division that follows is the anatomy of one calculation, not a fork into two theories.

The calculation itself is two lines. With the background responding at rest and the compliant channel responding along its own path, at the frequency it samples, the dispersion relation is $k^2 = \rho\,[\,C_g(\omega-kU)\,(\omega-kU)^2 + C_\ell\,\omega^2\,]$; its first-order expansion is
\[
f \;=\; \frac{C_g}{C} \;+\; \frac{\omega}{2}\,\frac{C_g'}{C}\,,
\]
the complete coefficient (Appendix~A; supplementary script). What follows fixes the physical picture, then takes the result apart.

\paragraph{The analogy, stated in full.} It is worth fixing the picture completely before the anatomy begins, because every later section leans on it. Take water, and let its inclusions be steam: bubbles of the same substance in a different state. Nothing in the system is  foreign to anything else: liquid and vapor are one material, the interface a change of state, and a sound wave in such a medium is water's own wave passing through water's own other phase. The pockets are soft (vapor yields where liquid barely does) so they hold nearly all of the medium's compliance; the wave slows accordingly, squeezes each pocket as it passes, and near the Minnaert frequency  the pockets ring, more heavily damped for vapor than for permanent gas, through evaporation and condensation at the interface, but ringing all the same. Shape oscillations (the spherical harmonics of the bubble surface) exist as well, but couple weakly at long wavelength; the breathing mode does the acoustic work. Now read the constructive model in parallel: the substrate is the water; the atom is  the steam bubble: a region of the same substance in a different state, soft where the background is stiff, holding the compliance that sets the refractive index, squeezed by every passing crest, resonant at its own frequencies. Light approaching and entering an atom is sound approaching and entering a steam bubble. The one structural element the model adds is circulation: the atom-region is organized as a vortex structure, and circulation, unlike a passive pocket, is transported with its carrier as a matter of classical vortex dynamics, in the ideal-fluid regime  the model assumes: the pattern travels because the pattern is motion. Circulation is what promotes the spherical harmonics from weakly coupled surface ripples to the load-bearing modes of the structure \cite{meucci2026cavity}, and it is what decides custody below. Everything else transfers unchanged.

With the picture fixed, carried versus severed response reduces to events. A compliant inclusion in a wave field does two things in every cycle: it \emph{stores} (the crest squeezes it, the trough lets it re-expand) and it \emph{delivers}, handing the stored response back to whatever carries the wave onward. Whether the wave inherits the inclusion's drift depends on where the delivery lands. A passive vapor pocket delivers at its interface into liquid that is standing still: whatever downstream displacement the stored phase earned while riding the bubble is surrendered at the hand-off, and the inclusion-drift closures of Appendix~A are the bookkeeping  of that surrender; slip at the interface is what makes it possible. A circulating structure cannot surrender without ceasing to be the structure it is: its response travels as its circulation travels, stored and re-emitted within the same moving pattern \cite{meucci2026cavity}, so every cycle hands the wave the full downstream displacement of its carrier. Storage in motion with delivery at rest loses the share; storage and delivery in motion keep it. Within mechanical inclusion accounts, Fizeau's measured coefficient is the empirical statement that the atom-inclusions keep custody: the share appears in no analyzed configuration that surrenders it (Appendix~A). The first-order equivalence with exact composition is broken at second order, where the gas-mode observable of Appendix~B separates the two. The resonance term, by contrast, needs only the storing instant (a drifting inclusion is detuned the moment the crest reaches it, sampling the wave at its own Doppler-shifted frequency), which is why the derivative law survives both delivery behaviors while the share distinguishes them.
\subsection{The compliance language and the share identity}

Write every medium's linear response as a total compliance $C$ assembled from a background part and a soft-inclusion part. For light in matter, in units of $\varepsilon_0$: $C_{\mathrm{vac}} = 1$, $C_{\mathrm{matter}} = \chi = n^2-1$, $C = 1+\chi$. For sound in a bubbly liquid: $C_{\mathrm{gas}} = \alpha/K_g(\omega)$, $C_{\mathrm{liq}} = (1-\alpha)/K_\ell$, $C = C_{\mathrm{gas}}+C_{\mathrm{liq}}$ \cite{commander1989,prosperetti1977,minnaert1933}, with $v^2 = 1/(\rho_{\mathrm{eff}}C)$ the Wood speed. In this language the optical non-dispersive coefficient is algebraically the \emph{share}:
\[
1-\frac{1}{n^2} \;=\; \frac{\chi}{1+\chi} \;=\; \frac{C_{\mathrm{matter}}}{C_{\mathrm{matter}}+C_{\mathrm{vac}}},
\]
the compliant inclusions' fraction of the total compliance, with the vacuum itself as background compliance. The acoustic share is likewise $C_{\mathrm{gas}}/C$; near Minnaert resonance $K_g(\omega)$ plunges, the share approaches unity, and the medium's compressibility becomes almost entirely bubble-borne. Section~4 develops the physical reading and its vindication of Fresnel's own excess-density intuition; here the identity serves as vocabulary: both laws below are statements about $C_{\mathrm{soft}}$ and $C$.

\subsection{The resonance law: universal, and now derived acoustically}

\emph{Claim.} For a bubbly liquid at rest except for a uniform drift $U$ of the bubbles, the dispersive part of the entrainment coefficient is
\[
f_{\mathrm{res}}(\omega) \;=\; \frac{\omega}{2}\,\frac{1}{C}\,\frac{dC_g}{d\omega}\,,
\]
exactly, at first order in $U/v$, and independently of the two-fluid closure: the result is unchanged whether the bubbles' perturbation motion is slaved to the liquid or given its full translational dynamics (the factor-three added-mass slip), and whether or not the added-mass impulse is advected with the bubbles. The configuration is the acoustic image of Fizeau's: the drifting bubbles play the moving matter (the water of the optical experiment) while the liquid plays the aether and never moves. The mechanism is Doppler evaluation of the resonant response: drifting bubbles experience the wave at $\omega' = \omega - kU\cos\theta$, and the medium's compliance is accordingly $C_g(\omega')$. The derivation, with sanity checks, is Appendix~A.

The optical counterpart is not a conjecture but the classical dispersive drag term. With $n^2 = 1+\chi$,
\[
\frac{\omega}{n}\frac{dn}{d\omega} \;=\; \frac{\omega}{2}\,\frac{\chi'(\omega)}{1+\chi} \;=\; \frac{\omega}{2}\,\frac{1}{C}\,\frac{dC_{\mathrm{matter}}}{d\omega}\,,
\]
identical in form and prefactor. This is the term Lorentz derived in 1895 from the Doppler response of moving molecular oscillators \cite{lorentz1895}, the term Zeeman's experiments were designed to discriminate and found favored \cite{zeeman1915}, the term measured by ring-laser interferometry \cite{bilger1972}, and the term that dominates modern anomalous-drag experiments in highly dispersive media, where group indices exceed $10^6$ and the drag is effectively all resonance law \cite{boyd2022}. One law, two domains, eleven decades apart: \emph{moving resonant compliance imprints its frequency-derivative on wave transport}, in Galilean media and Lorentzian media alike.

Both domains, moreover, exhibit the same normalized resonance structure: near resonance $(1/X)(dX/d\omega) \propto (-2\delta)/(w^2+\delta^2)$, where $X$ is $K_g(\omega)$ or $n(\omega)$, $\delta$ the detuning, $w$ the linewidth: the Lorentzian derivative form at kHz and at $10^{15}$~Hz.

The law requires no proximity to resonance. Substituting the Minnaert response $C_g(\omega)=C_g(0)\,\omega_0^2/(\omega_0^2-\omega^2)$ gives, below resonance, $f_{\mathrm{res}} \approx (\omega/\omega_0)^2\,\beta$ with $\beta = C_g(0)/C$: the bubbles entrain the sound by their compliance share, weighted by the squared  frequency ratio: the smooth tail of normal dispersion. This is the regime in which optics itself operates: visible light in water or glass lies far below the electronic resonances of the ultraviolet, and $n(\omega)$ in the visible is that same low-frequency tail. Approaching resonance the lossless formula gives way to the damped response, $C_g(\omega) \propto (\omega_0^2-\omega^2-i\Gamma\omega)^{-1}$, whose real part traces the classic dispersion profile: rising below resonance, falling steeply across the linewidth, recovering above. Within the linewidth $\mathrm{Re}\,C_g' < 0$ and the entrainment is expected to reverse (sound pushed against the bubble drift, the mechanical counterpart of anomalous optical drag \cite{boyd2022}), with peak magnitude set by the damping; the theorem of Appendix~A is exact for real compliances, so the reversal through the linewidth is stated as its weak-loss continuation, the full complex-wavenumber treatment remaining open (Section~10). The frequency-weighting also settles, physically, what Section~3 recounts as the century's sharpest objection: a single liquid with a single bubble population entrains each frequency by a different amount. There is not one water per pitch; there is one water with a frequency-dependent response. Chromatic \emph{response} requires no chromatic \emph{substance}.

\subsection{The deformation law: why the share belongs to closure-bound cavities}

\emph{Claim.} In the same two-fluid system with the resonance switched off (compliances frequency-independent), the drift of the bubbles produces \emph{no} share-shaped drag. The first-order coefficient collapses to a compliance-blind residue of order the void fraction ($\alpha/2$ under the no-slip closure; $\approx 11\alpha/4$ with full slip and added mass; Appendix~A), in which the ratio $C_g/C_\ell$ cancels identically; the wider family of Appendix~A can restore compliance-sensitivity to the residue, but in no member does the residue take the share form. Physically, within these severed-response closures a uniformly drifting compliant phase is invisible to the homogenized wave equation: the bookkeeping delivers the stored response to the inertia carrier's frame, and phase convection anchors there with it. The complementary Galilean statement checks against known flow acoustics: liquid flowing past resting bubbles convects the wave at nearly the full flow velocity.

The optical coefficient's non-dispersive part therefore cannot arise as severed-response inclusion drift. It takes the share value $C_{\mathrm{soft}}/C$ exactly where velocity composition is Lorentzian: expanding the relativistic composition of $c/n$ with $U$ gives $1-1/n^2$ \cite{vonlaue1907}, which Section~2.1 identifies as the share. We retain for the share term the name \emph{deformation law}; the deformation meant here is the closure-forced one, distinct from the ordinary hydrodynamic flattening any moving body suffers (Section~8). The name marks what holds in real media: a structure made of the medium's own waves cannot sever its response from itself, so its compliance is carried by construction and the share is guaranteed; and the same closure-forced deformation, carried to second order, generates exact velocity composition (the companion theorem), completing the law beyond first order. The share itself, at first order, asks only that the response be carried, and it is absent from every closure in the Galilean family analyzed in Appendix~A (a two-parameter class of slip and added-mass closures, further varied by a volume-coupled impulse) whose residues change with the closure but never once take the share form $C_g/C$; nothing in that family carries the response. The companion paper supplies the mechanism: cavities whose stability is phase closure of the medium's waves must, on pain of ceasing to exist, deform and retard so that their mutual measurements compose Lorentzianly \cite{meucci2026cavity}. Stated compactly: \emph{a cavity whose stationary geometry is fixed by phase closure of an internal mode of speed $c$ must, to preserve closure while translating at $U$, contract along the motion by $\sqrt{1-U^2/c^2}$, dilate its period by the inverse factor, and acquire the first-order phase  offset $Ux/c^2$, the local time} \cite{meucci2026cavity}. The present paper uses only these three outputs. The operative distinction must be drawn with care, because the structural analogy between the two cavities is tight, and it would be wrong to say one is an object and the other mere behavior. Both are objects, and for the same reason: a boundary in a medium adds a degree of freedom. A surface partially unlocks interior from exterior, which is how a cavity or a vortex can be an entity rather than an  episode of the flow: the walking-droplet systems display the same object-making at laboratory scale \cite{bush2015}. Both have an interior agency holding the cavity open (gas pressure in the bubble; circulation in a vortex-cored cavity) and an exterior confinement pressing it closed (surface tension in the one, the ambient stiffness of the medium in the other); and neither set of forces is instantaneous or wholly independent of the medium. What decides the deformation law is narrower: \emph{what fixes the cavity's geometry}. The bubble's radius is set by a static balance of pressures, indifferent to the phase of any sound wave; its Minnaert frequency follows from its size, not the reverse, so the wave field proposes nothing about its shape, and in motion it deforms only by flow forces, on flow scales. The closure-bound cavity's geometry is itself a solution of the wave-closure condition (the standing pattern, formed in the locally varying wave speed of the softened region, fixes the size), so a moving cavity must renegotiate its geometry with wave propagation, and the companion paper proves the renegotiation is uniquely the Lorentz deformation \cite{meucci2026cavity}. Geometry fixed by force balance: the response can be severed by slip, and in the drifting-bubble  configuration it is: resonance law only. Geometry fixed by wave closure: the response is  carried by construction: resonance law and share both, the deformation completing the share beyond first order. The bubble (its geometry carrying no closure constraint) realizes the compliance channel in isolation, and thereby tells the two laws apart. Table~\ref{tab:twolaws} summarizes the anatomy.

\begin{table*}[t]
\centering
\caption{The two laws inside the drag coefficient: form, realization in each domain, and epistemic status.}
\label{tab:twolaws}
\footnotesize
\begin{tabular}{l|l|l}
 & \textbf{Deformation law} $f_{0}$ & \textbf{Resonance law} $f_{\mathrm{res}}$ \\ \hline
Form & $C_{\mathrm{soft}}/C$ (share) & $(\omega/2)\,C_{\mathrm{soft}}'/C$ \\
Light & $1-1/n^2$ \cite{vonlaue1907} & $(\omega/n)\,dn/d\omega$ \cite{lorentz1895,zeeman1915} \\
Sound (drifting bubbles) & absent: $O(\alpha)$, never share-shaped & $(\omega/2)\,C_g'/C$ exactly \\
Status & signature of closure-forced deformation & universal; derived and measured \\
\end{tabular}
\end{table*}

\subsection{Group velocity, energy transport, and what kinematics alone omits}
The dispersive term $(\omega/n)(dn/d\omega)$ in the full drag coefficient (absent from constant-index velocity composition) reveals content the composition rule alone does not supply.

The full coefficient can be rewritten through group velocity:
\[
f_{\mathrm{opt}}(\omega) = \frac{n_g(\omega)}{n(\omega)} - \frac{1}{n^2(\omega)},
\]
where $n_g \equiv n + \omega(dn/d\omega)$ is the group index governing, in transparent regimes, the energy-transport velocity $v_g = c/n_g$, distinct from phase velocity $v_p = c/n$.

This distinction is fundamentally mechanical: away from resonant absorption, group velocity is the observable that tracks energy transport, distinct from kinematic phase tracking. 
Near resonances in bubbly liquids or atomic media, $v_g$ can differ from $v_p$ by factors of 2--3, and near strong anomalous dispersion $v_g$ famously becomes  negative or superluminal, precisely because it there ceases to track energy transport, which remains subluminal and forward; the taxonomy of group, signal, front, and energy velocities is Brillouin's \cite{brillouin1960}. 
These are energy storage and release effects with no counterpart in pure velocity-addition kinematics, which tracks only phase velocity.

Constant-index velocity composition reproduces only the non-dispersive term $(1-1/n^2)$, corresponding to phase velocity. The dispersive term enters, under any kinematics, only when the medium's resonant response is sampled at the Doppler frequency of its rest frame (Section~7); its form is a signature of resonant compliance, not of the composition rule. Mechanical models (bubbles, polarizable atoms) generate both $v_p$ and $v_g$ through that resonant dynamics.

Both acoustic and optical dispersion near resonances follow identical normalized form: $(1/X)(dX/d\omega) \propto (-2\delta)/(w^2+\delta^2)$ where $X$ represents either $K_g(\omega)$ or $n(\omega)$, $\delta$ is normalized detuning, and $w$ is linewidth.
This Lorentzian derivative structure appears in bubble resonances (kHz) and atomic transitions (10$^{15}$ Hz), eleven orders of magnitude apart.
The mathematical identity across scales suggests dispersion, whether acoustic or optical, arises from compliant resonators responding to wave fields.

\subsection{Three distinctions to hold}

The reader is now equipped against the three conflations on which, we will argue, a century of  interpretation foundered, and which Section~10 shows are still live.

\emph{Participation is not phase.} The rate at which the wave-motion (its energy, its disturbance) partakes of a component's motion is a residence quantity; the velocity of the interference pattern is a composition quantity. Veltmann's 1870 vocabulary (``theilnimmt'') is participation language \cite{veltmann1870}; the post-1907 tradition recast it wholesale as phase composition, legitimately for Lorentzian media, where the two coincide at first order, and misleadingly everywhere else.

\emph{The resonance term is not a correction.} The constant-index framing treats $1-1/n^2$ as the main term and the derivative term as a refinement. Near resonance the hierarchy inverts, in bubbly liquids and in slow-light media alike \cite{boyd2022}; and it is the derivative term, not the ``main'' term, that is universal mechanics.

\emph{The law is not the ontology.} Fizeau severed them in 1851 \cite{fizeau1851}; the multiple-ethers objection binds them back together by reading the coefficient as a bodily quantity of dragged medium. Every argument in this paper is about the law; the ontological claim is  sufficiency, not uniqueness: the mechanics generates the law, and whether nature has another way is not adjudicated here. A reader who takes the acoustic system as formal analogy alone forfeits nothing: the two-law separation, the closure-family theorems, and the documentary findings stand as results about the analog system and the historical record, whatever further reading is withheld.

\medskip
\noindent This completes the claim. Sections~2.1--2.3 carry the paper's load (one identity, one derivation, one negative result) and nothing after this point is required in order to believe them. What follows adds, in order: the history that fused the laws, the physical picture that explains their shapes, the contested episodes re-examined with the laws in hand, and the subtleties handled explicitly as subtleties. The simple picture stands on its own; the complexities are additions to it, not conditions on it.

\section{The interpretive history, 1818--1915}

A note on method. The historiographical claims of this section rest, wherever possible, on the actors' own words (Fizeau's severance of law from hypothesis in the founding report, Veltmann's participation vocabulary, Michelson and Morley's italicized conclusion), quoted with pinpoint citations. The analytical vocabulary of Section~2 is used to \emph{read} the record, not to grade it: where a framing is retrospective, as with the fork of Section~3.2 and the counterfactual of Section~9, the text says so at the point of use. In this the section follows the documentary standards of the modern historiography of the optics of moving bodies \cite{darrigol2000,janssenstachel2004,stachel2005}.

\subsection{Fresnel's excess-density hypothesis}

Fresnel's 1818 proposal, made in response to Arago's null result on stellar refraction, was that only the \emph{excess density} associated with transparent matter drags the luminiferous medium, which led directly to his $(1-1/n^2)$ coefficient \cite{fresnel1818}. The episode is documented in detail by historians of the optics of moving bodies \cite{darrigol2000,stachel2005}. The nuance in Fresnel's hypothesis is illuminating and easily lost: it was already a model of \emph{differential response}. The part of the medium that differs from vacuum (the ``inclusion'') does the dragging. In this respect Fresnel's reasoning aligns naturally with the compliant-inclusion framework developed below, in which the zones that differ from the background medium determine the optical effect. What Fresnel fixed, and what hardened into orthodoxy, was the \emph{kind} of difference: excess density. The choice is visible in the letter's own mathematics: Fresnel obtains the interior aether's density from the wave-speed relation with the elasticity held equal, $D'/D = v^2/v'^2$, so that the entire optical difference is booked as density \cite{fresnel1818}. The fork of the next subsection is thus textual at the level of the assumption made, even where the alternative branch was not yet articulable.

\subsection{The rigidity path not taken}

Wave speed in an elastic medium is $v=\sqrt{\mu/\rho}$. Reducing wave speed can arise from increased density ($\rho\uparrow$) or decreased rigidity ($\mu\downarrow$). The two options are mathematically equivalent at the level of $v$; they are ontologically distinct, and they entail different mechanics of drag.

This fork is  partly retrospective, and that is itself part of the story. It is not wholly retrospective, however: in the note additionnelle appended to the 1818 letter, Fresnel fenced his own mechanism: ``whatever the hypothesis one makes concerning the causes of the slowing'' of light, one may substitute for the real medium an equivalent elastic fluid whose density is chosen to match each propagation speed, and on that substitution, he states, the calculation rests \cite[pp.~3--4]{stachel2005}. The declared basis is an equivalence indexed by propagation velocity; the calculation was explicitly insulated from any unique commitment about the physical cause. The correction made here (compliance where Fresnel wrote density) selects within the substitution space Fresnel himself declared open. In 1818 the rigidity branch was scarcely articulable: elastic-solid theories of the aether (Cauchy's and Green's compressional media, MacCullagh's rotationally elastic medium that resists twist rather than compression) were developments of the 1830s and 1840s \cite{whittaker1951,schaffner1972}. The principal mid-century rival to Fresnel was instead Stokes's fully entrained aether, contrived with irrotational flow to preserve stellar aberration \cite{stokes1845}. Fizeau's 1851 measurement returned the \emph{partial} coefficient, carrying Fresnel's hypothesis to experimental authority over Stokes's \cite{fizeau1851}. Fizeau himself pried the law from the ontology at the moment of confirmation: the experiment, he wrote, entails the adoption of Fresnel's hypothesis, ``or at least of the law that he found,'' while the conception itself would appear so extraordinary as to require still more proofs and a deepened examination before acceptance (\cite{fizeau1851}; translation from \cite[p.~6]{stachel2005}). The law was severed from its mechanism by the coefficient's  own confirmer, and yet the density reading traveled onward with the law.

The density reading then generated its own crisis, and the crisis is diagnostic. Because $n$ depends on color, the excess density $\propto n^2 - 1$ does too: taken literally, transparent matter must contain, and drag, a different \emph{quantity} of aether for every frequency. Wilhelm Veltmann derived that requirement in 1870, from the standpoint of relative motion, with the coefficient emerging as necessary rather than assumed \cite[cols.~145--146]{veltmann1870}, and he drew the narrow verdict, not the broad one. The participation velocity, he states, depends on the propagation velocity and so must differ for each color, and in birefringent media for the two polarizations; the hypothesis had accordingly been put forward ``nur als eine mathematische Beziehung'' (only as a mathematical relation), his talk of aether motion ``only for brevity of expression'' \cite[cols.~159--160]{veltmann1870}. Fresnel's particular picture of a bodily entrained excess could not carry values that multiply with color and polarization; mechanism itself was another matter. His closing sentence files that verdict explicitly: ``Die wirkliche physikalische Erkl\"arung der Aberrationserscheinungen geh\"ort jedenfalls zu den schwierigsten Aufgaben der Vibrationstheorie'': the actual physical explanation of the aberration phenomena belongs, in any case, among the hardest tasks of the wave theory \cite[col.~160]{veltmann1870}. By 1873 the relation stood as the necessary and sufficient condition for carrying rest-media optics over to moving media, its physical foundation set aside as worthless for that purpose \cite[pp.~6--8]{stachel2005}; Lorentz adopted the ray-based method by name in the 1895 \emph{Versuch} \cite{lorentz1895}, and a 2026 study in \emph{Annalen der Physik} credits Veltmann as instrumental in the birth of nineteenth-century ``optical relativity'' \cite{bracco2026}. In transmission, however, the narrow verdict broadened: read against a literal bodily aether, the chromatic requirement became the famous reductio (one glass obliged to drag a different aether at a different speed for every color \cite[p.~7]{stachel2005}), and hardened in reception toward the broader conclusion that no mechanical account could do. The primary sources support only the narrower verdict, and the narrower verdict names an assignment this paper completes: the per-color participation is the resonance law of Section~2.2, the medium's tuned compliance sampled at each color's Doppler-shifted frequency (the same tuned response that makes the index itself frequency-dependent at rest) and the per-polarization participation is anisotropic compliance, two response values along two axes. The dependence is exactly as innocent as refraction's own: one prism, one law, a different angle for every color, and no one infers a separate glass for each. Under the compliance reading there is nothing to object to: a resonant compliance is chromatic by nature, and the frequency-dependence of its share \emph{is} the coefficient's dispersion, not a paradox.

The nineteenth century in fact resolved the crisis in  precisely this direction: by keeping the law and quietly abandoning the ontology. Historians document a growing pattern, from Veltmann and Mascart through Fizeau himself to Lorentz and Poincar\'e, of affirming the coefficient's empirical necessity while disclaiming its literal drag interpretation \cite{janssenstachel2004}. And when Michelson and Morley repeated Fizeau's experiment with improved precision in 1886, they confirmed the coefficient and drew the ontological moral themselves, concluding that the luminiferous aether is ``entirely unaffected'' by the motion of the matter it permeates \cite[p.~386]{michelson1886}. The same experiment that certified the law rejected the bodily drag. Section~2.2's tank now performs the dissolution physically: one liquid, one bubble population, a different entrainment  for every pitch, and not one water per pitch. What no one then supplied was a replacement ontology in which a frequency-dependent \emph{share} is natural; Section~4 shows the coefficient is exactly such a share, and that Fresnel had it from the beginning.

The rigidity alternative in its mature form holds that atoms are regions of \emph{reduced rigidity} within a continuous substrate: light slows because the medium is locally softer, not denser. This reading naturally accommodates  topological structures: vortex configurations that maintain stability through circulation, of which quantized vortices in superfluid helium-II are the modern laboratory exemplar \cite{donnelly1991}. When Kelvin introduced vortex atoms in 1867 \cite{kelvin1867}, his ontology supplied precisely the kind of structure (persistent through the flow of its own medium) that a rigidity reading would, in retrospect, need in order to unify refraction and partial drag. But by then Fresnel's coefficient carried Fizeau's authority under the density interpretation, and after 1865 the community's theoretical energies flowed to electromagnetic theory; the vortex-atom program pursued spectra and the constitution of matter rather than a rework of the optics of moving bodies. The rigidity path remained unexplored in mainstream development.

\subsection{Michelson--Morley and the charge of special pleading}

Fresnel's coefficient compensated first-order optical effects while leaving the bulk aether essentially stationary; the Earth's residual motion through that stationary background remained, and second-order experiments were designed to reveal it. Michelson and Morley found nothing \cite{michelson1887}. Heaviside, meanwhile, had computed that the field of a charge in uniform motion through the electromagnetic medium is compressed into oblate spheroidal form with aspect ratio $\sqrt{1-v^2/c^2}$ \cite{heaviside1892}: Lorentzian geometry obtained from the structure of moving systems in a medium, years before any spacetime postulate. FitzGerald, who was in correspondence with Heaviside on moving-charge fields \cite{fitzgerald1889}, and Lorentz independently in 1892 \cite{lorentz1892} (developing the hypothesis within his electron theory through to the 1904 transformations \cite{lorentz1904}) responded to the null result with the deformation hypothesis: material bodies deform at second order by just enough to  null the interferometer: at first any member of a one-parameter family of longitudinal--transverse deformations compatible with the null, only later fixed to the purely longitudinal $1/\gamma$ \cite{brown2005}; the closure theorem of the companion is what singles that member out uniquely \cite{meucci2026cavity}. Contemporary critics (most influentially Poincar\'e) objected to the resulting structure of the theory: exact compensations, introduced one by one, each tailored to the experiment it explained \cite{darrigol2000}. Two such compensations stood out. Fresnel's coefficient was the first-order compensation: it explained why no optical experiment revealed the Earth's motion  at order $U/c$, and it did so as prediction, thirty-three years before Fizeau measured the corresponding moving-medium effect. The FitzGerald-Lorentz contraction was the second-order compensation, nulling the Michelson--Morley signal at order $U^2/c^2$. Two independent fixes appeared to be special pleading. We return to this charge in Section~9, where the mechanical framework shows the two compensations to be one.

\subsection{Von Laue's derivation and the elimination verdict}

The coefficient's empirical authority reached into the foundations of the new kinematics itself: Einstein, interviewed late in life by Shankland, recalled that the experimental results which had influenced him most were stellar aberration and Fizeau's measurement: ``they were enough'' \cite{shankland1963}. In 1907 von Laue showed that Fresnel's coefficient follows, to first order, from the relativistic composition of velocities \cite{vonlaue1907}. The derivation is short, exact at its order, and requires no medium. It is, moreover, mathematically equivalent to the derivation Lorentz had already given in 1895 within an explicitly medium-based theory \cite{lorentz1895,vonlaue1907}: the same coefficient, derived in two ontologies, of which the tradition retained one. What von Laue showed, precisely, is that the coefficient no longer required a separately postulated drag mechanism: it follows from velocity composition alone. In the canonical relativistic literature this hardened into a stronger reading (partial drag as ``just kinematics,'' the mechanical question dissolved rather than answered), a reception the historiography itself documents \cite{janssenstachel2004}. Pauli's authoritative 1921 review is representative: the coefficient appears there as a consequence of the composition law, with no mechanical hypothesis remaining in play \cite{pauli1921}. The kinematical response was, moreover, latently available well before 1905: Mascart had formulated an optical principle of relative motion by 1874 (``relative motions are the only ones we are able to determine'') and Stachel himself asks whether a kinematical resolution might have been developed from it around 1880 \cite[pp.~8,~10]{stachel2005}.

Yet what a kinematic derivation shows about underlying dynamics has never been settled. Bell argued that the ``Lorentzian pedagogy'' (deriving relativistic effects from the dynamics of matter in a preferred frame) remains legitimate and instructive \cite{bell1976}; Brown has developed a book-length case that length contraction and time dilation are ultimately dynamical phenomena \cite{brown2005}; Janssen has argued, against this, that the kinematic reading is explanatorily superior \cite{janssen2009}. The present paper does not adjudicate that debate. It contributes something the debate has lacked: a laboratory transport system in which the kinematic derivation and the mechanical ground are \emph{simultaneously} available and demonstrably compatible. Seen from this vantage, the constructive tradition of FitzGerald, Heaviside, and Lorentz was never refuted in 1907; it was set aside with its central uniqueness question unposed. The companion paper poses and answers that question \cite{meucci2026cavity}; the present paper recovers the transport-side history that made the question invisible.

\section{The physical principle: waves ride the soft component}

With the two laws established, this section and the next widen the physical picture behind them. Section~2 stated the principle (waves couple to the most compliant component) and the anatomy proved it; the question here is why the principle recurs with the same shape across domains. The answer is mechanical: \emph{wave energy couples to the most compliant part of the medium}, the subsystem most deformable by the propagating field.

The identity can be made exact rather than analogical. Write the optical coefficient in terms of the susceptibility $\chi = n^2 - 1$:
\[
f_{\mathrm{opt}}^{(0)} \;=\; 1 - \frac{1}{n^2} \;=\; \frac{\chi}{1+\chi} \;=\; \frac{C_{\mathrm{matter}}}{C_{\mathrm{matter}} + C_{\mathrm{vac}}},
\]
where $C_{\mathrm{vac}} = \varepsilon_0$ is the vacuum's baseline electromagnetic compliance and $C_{\mathrm{matter}} = \varepsilon_0\chi$ is the additional compliance contributed by the inclusions. That the vacuum itself carries electromagnetic response is consonant with established quantum electrodynamics: the Heisenberg--Euler vacuum acquires field-dependent polarization \cite{heisenberg1936}, and strong observational evidence for the resulting birefringence has now emerged from magnetar X-ray polarimetry \cite{stewart2026}. Three claims must be kept distinct: the identity above is algebra; reading $\varepsilon_0$ and $\varepsilon_0\chi$ as baseline and added constitutive compliance is standard electrodynamics renamed; reading that compliance as the deformability of a mechanical substrate is the constructive model  this paper advances, as sufficient, not unique. Set this beside the acoustic coefficient,
\[
f_{\mathrm{bub}} \;=\; \frac{C_{\mathrm{gas}}}{C_{\mathrm{gas}} + C_{\mathrm{liq}}},
\qquad C_{\mathrm{gas}} = \frac{\alpha}{K_g}, \quad C_{\mathrm{liq}} = \frac{1-\alpha}{K_\ell},
\]
and the two are one expression: \emph{the entrainment coefficient is the compliant inclusions' share of the total compliance.} Fresnel's non-dispersive coefficient is not analogous to the acoustic fraction; it \emph{is} that fraction, with the vacuum playing the liquid's role as background compliance. And Fresnel's own excess-density hypothesis (only the part of the medium that differs from vacuum is dragged) was this share law from the beginning, stated in the wrong ontology: replace excess density with added compliance and the 1818 intuition becomes exact.

The certification is asymmetric, and the asymmetry is the finding. Optics certifies the deformation law: relativity and a century of interferometry fix the share value of the non-dispersive term, which Section~2.3 shows no member of the analyzed Galilean inclusion-drift family supplies. Acoustics certifies the resonance law: the derivation of Appendix~A fixes the derivative term exactly, closure-independently, in a system whose mechanics is fully transparent. Each domain proves the law the other cannot.

\textbf{In bubbles}, this is directly observable. 
Acoustic pressure oscillates bubbles (compressible) far more readily than liquid (incompressible). 
Wave energy resides predominantly in bubble oscillations. 
When the bubbles drift with velocity $U$, the wave is entrained through their resonant response (the resonance law of Section~2.2) and the wave's energy, which at Minnaert resonance resides wholly in the bubble's oscillation mode (gas compression exchanging with the co-moving near-field inertia of the surrounding liquid) participates in their motion in Veltmann's sense (Section~10).

\textbf{In atoms}, optical fields shake electron clouds (polarizable) while nuclei remain essentially rigid. 
The electron cloud plays the role of the bubble: the soft, responsive component. 
Cloud size and compliance determine both refractive index $n(\omega)$ and drag coefficient $f_{\mathrm{opt}}(\omega)$.

For atoms, the resonance frequency scales with electron cloud radius $a$:
\[
\omega_{0} \sim \sqrt{\frac{e^2}{4\pi\varepsilon_0 m a^3}}, \quad \lambda_0 \propto a^{3/2} m^{1/2}.
\]
Polarizability (cloud size) dominates; mass enters weakly.

\textbf{In relativity}, first-order velocity addition encodes the deformation law abstractly. 
The factor $(1-v^2/c^2)$ in $f_{\mathrm{SR}}$ determines entrainment strength, with $v(\omega)$ playing the role of ``effective compliance'': slower wave speed means greater entrainment.

Bubble dynamics operates at the scale of gas-liquid mixtures, far above vacuum  substrate dynamics, just as planetary motion operates above atomic structure without invoking quantum mechanics at each instant. 
But the principle is scale-independent: the resonant transport channel couples to the compliant, wave-accessible component of the medium.

\section{Extension: polarization as mechanical deformation}

The bubble-atom analogy extends  beyond wave propagation: it reveals the mechanical basis of atomic polarization itself.

Bubbles respond to acoustic fields in two modes. 
In the long-wavelength regime $\lambda \gg R_0$, the incident pressure is uniform across the bubble at leading order, driving \emph{monopole (breathing) oscillations}: the bubble expands and contracts radially, determining sound speed and refraction. 
At the next order in $kR_0$, the bubble samples the \emph{pressure gradient} across itself, and responds by translating bodily:
\[
\mathbf{x}_{\text{bubble}} \propto -\nabla P.
\]
This is \emph{dipole oscillation}: the bubble and its flow field form an acoustic dipole with strength $\propto R_0^3$.

Atomic polarization follows identically. 
An atom in electric field $\mathbf{E}$ exhibits charge separation: the electron cloud (soft) shifts by $\boldsymbol{\delta}$ while the nucleus (rigid) remains fixed, creating dipole moment $\mathbf{p} = \alpha \mathbf{E}$ where polarizability $\alpha \sim 4\pi\varepsilon_0 a^3$ scales with cloud volume.

Both systems obey the same driven-oscillator form, $m_{\mathrm{eff}}\ddot{\mathbf{x}} + k_{\mathrm{eff}}\,\mathbf{x} = \mathbf{F}$, forced by $\mathbf{F} \propto -\nabla P$ for the bubble and by $q\mathbf{E}$ for the electron cloud; term by term, the correspondence is exact at  the oscillator level: a mechanical isomorphism (Table~\ref{tab:mapping}).
\begin{table*}[t]
\centering
\caption{The bubble-dipole/atomic-polarization correspondence: a mechanical isomorphism at the oscillator level.}
\label{tab:mapping}
\begin{tabular}{l|l}
\textbf{Bubble dipole} & \textbf{Atomic polarization} \\ \hline
Pressure gradient $\nabla P$ & Electric field $\mathbf{E}$ \\
Gas (soft) & Electron cloud (soft) \\
Liquid (rigid) & Nucleus (rigid) \\
Displacement $\mathbf{x} \propto R_0^3$ & Polarizability $\alpha \propto a^3$ \\
\end{tabular}
\end{table*}

If atoms are treated as low-rigidity zones in a structured medium, polarization is simply mechanical deformation of compliant regions  under applied fields, exactly as bubbles translate under pressure gradients.

Macroscopically, individual atomic polarizability $\alpha$ determines refractive index via Clausius-Mossotti:
\[
\frac{n^2-1}{n^2+2} = \frac{N\alpha}{3\varepsilon_0},
\]
where $N$ is atom number density \cite{jackson1999}. 
Similarly, bubble compliance determines sound speed through effective bulk modulus. 
Both encode the same principle: wave speed measures aggregate compliance of soft inclusions.

Isotope substitution supplies a discriminating datum. Suppose the refracting oscillator were the massive particle itself, as a naive mechanical reading of dispersion would have it: a response tied to that particle's inertia (define it by refractivity proportional to the massive particle's mean-square zero-point excursion, $n-1 \propto \langle x^2\rangle_0 = \hbar/(2m\omega_0) \propto m^{-1/2}$ at fixed stiffness) predicts $(n_{\mathrm{D}}-1)/(n_{\mathrm{H}}-1) = \sqrt{18/20} = 0.948$ for H$_2$O$\to$D$_2$O, a 5.2\% isotope effect; an electronic-compliance response at fixed stiffness, $\alpha(0)=q^2/k$, predicts essentially no molecular-mass dependence at all. The measured effect is $\approx$1.4\% \cite{kedenburg2012}, corresponding to an effective mass exponent $\approx 0.13$ rather than $0.5$; the H$_2$/D$_2$ pair (measured at microwave frequency, in the static-polarizability regime where $\alpha(0)=q^2/k$ applies directly) gives $\lesssim 0.04$ \cite{essen1953}, the mass-independence of the static compliance observed as such. Molar-volume differences between the isotopologues are at the few-tenths-percent level and cannot close the gap. The data thus reject the inertia-coupled model as defined and sit where compliance-coupling puts them: the oscillating degree of freedom is the electronic structure, whose stiffness and extent are nearly  independent of nuclear mass: the conclusion of the standard electronic-oscillator picture, and equally the compliant-inclusion lesson. It is the acoustic result transposed: bubble response is governed by gas compressibility, not gas density \cite{minnaert1933}. Within the compliance reading, weak mass dependence is not an anomaly demanding explanation but the expected signature of soft-component coupling.

\textbf{In this reading: refractive index = measure of polarizability = measure of compliance.} The unification with Section~2 is then a single sentence: one softening, read in two circumstances. At rest, the softened cavities slow the wave and give $n$; in motion, closure forces the same cavities to deform and retard, and the same response $\chi(\omega)$ then yields the share $1-1/n^2$. The index is compliance read at rest; the deformation law is compliance read through motion.

Noble gases with larger electron clouds have higher refractive indices not as arbitrary correlation but as direct consequence: larger compliant zones respond more strongly, exactly as larger bubbles dominate acoustic response.

On this reading, the traditional mystery (why do charges separate under applied fields?) dissolves. 
Polarization is deformation. 
Compliant structures displace under gradients. 
Bubbles make this obvious; low-rigidity atomic models follow the same mechanics.

The frequency-dependent response producing dispersion (and thus group velocity distinct from phase velocity) follows naturally from resonant oscillators where $dn/d\omega \neq 0$. In modern terms the store-and-release has a name: excitation and decay. The medium absorbs into the cavity's internal mode, holds the energy for a lifetime, and re-emits (downstream of the absorption point when the cavity is moving) and the classical Lorentz oscillator is the smooth-limit bookkeeping of exactly this. Operationally the mechanism stands on its own (it is what makes the mathematics compute and the physics teachable) and Section~10 states what more the construction establishes. 
The rigidity-based interpretation connects to acoustic metamaterials where negative index arises from engineered density and modulus \cite{nature_acoustic_negative}: anomalous dispersion ($dn/d\omega < 0$) corresponds to medium stiffening ($dK_{\mathrm{eff}}/d\omega > 0$), directly observable in bubbly liquids where $K_g(\omega)$ minimizes at resonance then rises. 
This cross-domain framework suggests engineering techniques for controlling dispersive response may transfer between acoustic and optical domains.

\section{Vortices versus bubbles: directional asymmetry}

The historical record of Section~3 poses a mechanical question the framework must answer: if matter drags a medium, why did Michelson and Morley see nothing? The answer turns on a distinction invisible in Eq.~(\ref{eq:template}) but critical mechanically.

Bubbles and vortices differ crucially. 
\textbf{Bubbles move \emph{with} the liquid}: they are mechanically entrained. 
Push the liquid and bubbles flow along. 
\textbf{Vortex structures have circulation patterns through which the medium flows.} 
A vortex ring translates while fluid streams through its core.

If atoms mechanically drag substrate like bubbles drag liquid (the natural development of Fresnel's density reading), Michelson--Morley should have seen strong signals. 
But if atoms are vortex structures in the sense of Kelvin \cite{kelvin1867}, substrate flows \emph{through} them largely unimpeded. 
The optical effect (partial drag coefficient $f_{\mathrm{opt}}$) arises from wave propagation through reduced-rigidity zones, not bulk convection of substrate.

The null result becomes compatible with structured-medium models when this mechanical distinction is recognized. Historically, the distinction was available from 1867 onward; it was never brought to bear on the drag problem.

\section{Reversing von Laue: derivability is not elimination}

Von Laue showed that Fresnel drag follows from relativistic velocity addition \cite{vonlaue1907}, and the result was received as eliminating mechanical explanations: partial drag became ``just kinematics.'' In the two-law anatomy of Section~2 the reception's error becomes precise: von Laue kinematized the \emph{deformation} law (compressed its value into velocity composition) while the \emph{resonance} law Lorentz had already mechanized in 1895, and it remains mechanical in every medium. One part's derivability was taken for the whole's dissolution.

The verdict fails on two independent grounds, each worth stating as sharply as the elimination was.

\emph{First ground: the kinematics is a derivation route,  not a source, for either term.} That the drag admits a physical, Doppler-based derivation with no essential reliance on special relativity was shown by Drezet \cite{drezet2005}; the carried-response calculation gives that argument constitutive form and extends it to the complete coefficient. Velocity composition alone yields the non-dispersive share; the dispersive term enters only when the medium's response is evaluated at the Doppler frequency  of its rest frame: that is, only when a resonant constitutive response is imported into the derivation. Appendix~A shows the same term arises, with the same prefactor, in a Galilean medium: the resonance law holds under Galilean and Lorentzian kinematics alike, and is therefore a signature of resonant compliance, not of relativistic kinematics. What is distinctively relativistic in the coefficient is the share value alone.

\emph{Second ground: the law that was read as kinematics is itself a theorem of mechanics.} Lorentzian composition is not a primitive that terminates explanation; the companion paper derives it as the closure condition for resonant cavities of the medium itself \cite{meucci2026cavity}, and the bubble, whose geometry owes nothing to acoustic closure, obediently fails to exhibit the share, exactly as that derivation requires. The chronology already pointed this way: Lorentz obtained the coefficient in 1895 from a dynamical theory, with local time as bookkeeping \cite{lorentz1895}; von Laue's 1907 derivation re-derived from postulates what dynamics had produced twelve years earlier \cite{vonlaue1907}. The kinematic derivation did not supply a missing mechanism; it compressed one.

So the verdict is overturned in full. ``Pure kinematics, inexplicable by mechanics'' becomes: one law that is mechanics under every kinematics tested, and one law whose  kinematics is mechanics: the mechanics of measurement among closure-bound cavities. Derivability from kinematics eliminated nothing; it located the second place where the mechanics lives. The reception embeds an inference: \emph{if a phenomenon is derivable from kinematics alone, mechanical accounts of it are superfluous}.

The acoustic system is a counterexample  to the inference, not to the derivation. In a bubbly liquid, long-wavelength perturbations obey an effective kinematics from which the first-order transport law, Eq.~(\ref{eq:template}), can be derived without any reference to bubbles; this is the standard situation in acoustic analogues, where quasiparticle propagation is governed by an effective metric \cite{unruh1981,barcelo2011}. Yet in the same system the mechanism (the resonant compliance of the gas phase, Doppler-sampled by the drifting bubbles) is independently known, experimentally manipulable, and now derived to be quantitatively responsible for the resonance part of the coefficient (Appendix~A). Derivability from an effective kinematics is therefore compatible with, indeed generated by, an underlying mechanics.

\textbf{If explicit mechanics generates the same first-order law, then derivability of that law cannot certify the absence of mechanics.} Von Laue's theorem proves that the optical coefficient follows from Lorentz kinematics; it cannot prove that no mechanics underlies Lorentz kinematics. The stronger reading (that the relativistic law preserves mechanical content in abstract form) is then not merely permissible but constructed: the resonance part is generated by compliant response (Appendix~A), the share part by closure-forced deformation through local time \cite{meucci2026cavity,lorentz1895}. Mechanical sufficiency, not ontological uniqueness, is the claim (a generating construction of the law at its order) and sufficiency is precisely what the received verdict denied. Von Laue abstracted \emph{away} from mechanics; the bubble-relativity equivalence permits abstraction \emph{toward} them. The factor $(1-v^2/c^2)$ in $f_{\mathrm{SR}}$ need not be viewed as brute kinematics, but can be read as the signature of wave propagation through compliant inclusions  in an elastic medium, similar to how thermodynamics successfully abstracts away from the mechanics of individual molecules while remaining grounded in statistical mechanics. The companion paper carries this to completion for metrology: instruments built from closure-preserving resonant structures generate the Lorentz transformation as the self-consistency condition of their own measurements \cite{meucci2026cavity}: an explicit mechanism by which a law can be exactly derivable from kinematics while being generated by mechanics.

The interlock is exact at the level of orders. The closure metrology has three ingredients: contracted rulers, dilated clocks, and the first-order synchronization offset $-Ux/c^2$: Lorentz's \emph{local time}. It was from local time that Lorentz derived the Fresnel coefficient in 1895 \cite{lorentz1895}. The lineage runs earlier still: Potier had derived the $lu/V^2$ time offset directly from Fresnel's formula in 1874, an expression Stachel identifies as formally the local time \cite[pp.~8--9]{stachel2005}. Partial drag is therefore the first-order face of the same closure structure whose second-order faces are contraction and dilation: one transformation, read at successive orders in $U/c$, yields this paper's transport law and the two ``compensations'' of Section~9 alike.

This position has a live scholarly context on both flanks. Within the foundations of relativity, Bell's Lorentzian pedagogy \cite{bell1976} and Brown's dynamical analysis \cite{brown2005} argue that kinematic derivations rest on dynamical facts; Janssen replies that the kinematic reading carries the explanatory weight \cite{janssen2009}. Within the philosophy of analogue experiments, the question of when an analogue system supports inference about its target (developed precisely for acoustic analogues of gravity) is treated as open and analyzable rather than dismissed \cite{dtw2017}. We situate our claim at the intersection, and no further: the resonance-law correspondence is exact at first order, with identical prefactor; the mechanism is explicit in the acoustic member; and declaring the correspondence ``mere analogy'' is itself a metaphysical commitment, exactly as much as declaring it identity. Either commitment remains available to the reader; the theorems of Section~2 and Appendix~A require neither. The historical record of Section~3 shows that the former commitment was adopted, not demonstrated.

\section{Contraction as deformation of compliant structures}

A moving bubble in liquid does not remain spherical. 
Dynamic pressure is highest at the fore and aft stagnation points and lowest around the equator; the bubble flattens along the direction of motion. 
The equilibrium form is an oblate spheroid, compressed along the motion.

This is not incidental but fundamental to partial-drag dynamics.

The angular dependence $\cos\theta$ in Eq.~(\ref{eq:template}) encodes directional asymmetry: wave speed differs parallel versus perpendicular to motion. 
By itself the angular factor is the generic geometry of convection; it does not, alone, establish compression. The compression of moving structures enters through the closure theorem below, with the bubble's hydrodynamic flattening (directly observable in flow) as its qualitative exhibit.

Relativistic kinematics exhibits the same Lorentzian structure twice, with different arguments. 
Evaluated on the wave speed $v(\omega)$, $1-v^2/c^2$ sets the entrainment strength $f_{\mathrm{SR}}$; evaluated on the body's speed $U$, its square root sets the contraction:
\[
L = L_0\sqrt{1 - U^2/c^2}.
\]
One concerns the wave, the other the structure; their kinship is theorem, not numerology: phase closure in a medium of invariant wave speed forces both \cite{meucci2026cavity}.

If atoms are modeled as low-rigidity zones producing both refraction and partial drag, they must  contract when moving, just as bubbles deform in flow. 
A ruler composed of such structures inherits this compression. Condensed-matter analogues make this more than metaphor: structures bound by phonon-mediated interactions in a medium undergo a real Lorentz-FitzGerald contraction governed by the sound speed \cite{barcelo2008}, the acoustic counterpart of Bell's Lorentzian pedagogy \cite{bell1976}.

The companion paper proves the corresponding uniqueness theorem: for a resonant structure moving through a mechanical wave medium, the Lorentzian aspect ratio $a_\parallel/a_\perp = 1/\gamma$ is the \emph{unique} deformation preserving spherical-harmonic phase closure, and the same closure condition yields the period dilation $T=\gamma T_0$ with no further assumption \cite{meucci2026cavity}. The bubble supplies the qualitative, hydrodynamic exhibit; the closure theorem supplies the exact law for wave-organized structures, with contraction and dilation as paired consequences of one constraint. The oblate spheroid thus recurs across the program: hydrodynamically in the moving bubble, electromagnetically in the field of Heaviside's moving charge \cite{heaviside1892}, and exactly in the closure theorem.

\textbf{Within the structured-medium model, Michelson and Morley's null result becomes explicable}: the apparatus contracted exactly enough to compensate the optical path difference. 
Bubble deformation, velocity addition, and length contraction are three aspects of one principle: compliant structures in flow deform along the direction of motion.

\section{The two compensations were one}

Section~3 recorded the charge of special pleading: two suspiciously convenient ``compensations,'' Fresnel's coefficient nulling the Earth's motion in optics at first order, and the Lorentz-FitzGerald contraction nulling it at second order \cite{lorentz1904,fitzgerald1889}. Two independent fixes appeared as two arbitrary hypotheses. The pattern had been named from inside the era: Veltmann opened his 1870 analysis complaining that Fresnel's hypothesis had so far served only one special observation or another, ``immer um eine sogenannte Compensation'': always by way of a so-called compensation \cite[col.~145]{veltmann1870}.

But they are not independent. Both emerge from one mechanism: compliant structures deforming in flow.

The partial drag coefficient $f(\omega)$ and the contraction factor $\sqrt{1-U^2/c^2}$ are not one formula; they are two outputs  of one constraint: wave propagation and phase closure in the same medium \cite{meucci2026cavity}. 
In bubbles, entrainment and hydrodynamic deformation are simultaneously  present and observable: the qualitative exhibit of the pairing; the exact Lorentzian deformation belongs to the closure-bound cavity, by the companion theorem \cite{meucci2026cavity}. 
In vortex-structured models, the same dynamics apply.

FitzGerald and Lorentz were not inventing arbitrary adjustments. 
They were intuiting dynamics of compliant zones without the full framework. 
Their insights seemed disconnected only because Fresnel's density interpretation had obscured the mechanical picture seventy years earlier.

Counterfactually: had the rigidity path been developed through Kelvin's vortex atoms, these effects would have been recognized as unified consequences. 
The ``coincidence'' that appeared to doom classical optical theory (two phenomena requiring two fixes) was never a coincidence. 
It was one mechanism observed twice.

\section{Discussion: how the conflations reproduce themselves}

The historical record of Section~3 is not merely past. Each conflation catalogued there re-arises naturally in modern analysis, and the two-law division of Section~2 is the prophylactic. We record four instances.

First, the participation/phase conflation. Because the relativistic case fuses residence and composition at first order, the fused vocabulary became standard, and mechanical claims of the form ``the wave rides the soft component'' are now routinely evaluated as phase-velocity claims. Evaluated so, they fail  in Galilean media, correctly and irrelevantly: the residence claim concerns where the wave's energy sits and travels, a quantity that is resonance-dominated and, at Minnaert resonance, carried wholly by the bubble's oscillation mode: gas compression exchanging with co-moving near-field liquid inertia. The first-order law of energy transport in the drifting-bubble system is a natural sequel to the present work; we flag it as open rather than assert its value.

Second, the constant-index conflation. A two-fluid analysis of the drifting-bubble system performed with frequency-independent compliance (the natural formalization of the constant-index ``main term'') returns the compliance-blind residue of Section~2.3 and appears to refute mechanical entrainment altogether. The refutation refutes only the amputated system: restoring the resonant frequency-dependence recovers the universal law exactly (Appendix~A). The amputated calculation is included as a cautionary demonstration. The constant-index framing is not a neutral simplification; it deletes the mechanism under examination.

Third, the law/ontology conflation. The multiple-ethers reductio, the era's sharpest weapon against Fresnel, is an argument about the \emph{quantity} of a bodily dragged medium, and quantity-talk presupposes the density ontology (Section~3). Under the compliance reading the chromatic structure of the coefficient is not a paradox to be explained away but the  resonance law itself: the very part of the coefficient we have shown to be domain-universal. The objection that helped discredit mechanical readings was, all along, pointing at their most mechanical and most universal content. Fourth, configuration-severance mistaken for verdict. The inclusion-drift closures of Appendix~A tie the constitutive response to phase-transport bookkeeping; within that family the share never appears, and the absence invites the conclusion that mechanics cannot supply it. The carried-response calculation shows the conclusion to be a property of the family, not of mechanics: restore the response to its carrier and the share returns, the resonance term unchanged throughout. The diagnostic deserves statement for reuse: before reading a transport coefficient's absence as impossibility, ask what the closure did with the frame in which the response is evaluated: where the compliant channel stores, and where it delivers.

The division has one further edge, which we state conditionally and without verdict on the data; nothing in Sections~2--9 depends on it, and the paper's claims are unchanged if the paragraph is set aside. Appendix~B derives, under first-order transport closure of the two laws, the second-order sensitivity of a gas-filled Michelson interferometer: the classical fringe shift suppressed by $(n^2-1)$, so that a speed $U_N$ inferred under the classical calibration corresponds to $U = U_N/\sqrt{n^2-1}$ (a factor of about $41$ in air), while the vacuum instrument nulls exactly. Exact Lorentz covariance nulls the gas instrument as well; the gas-mode residual is therefore the observable that discriminates the two laws at their derived orders from exact covariance. The question is narrower than ``isotropy versus anisotropy.'' Mainstream physics already employs a non-constant coordinate speed of light (the Shapiro delay \cite{shapiro1964}) and a physically distinguished cosmic rest frame, against which the CMB dipole measures the solar system's motion \cite{planck2020}. What exact local covariance forbids is specifically a laboratory second-order signal in a co-moving gas; the dispute concerns only whether local gas-mode instruments couple to the motion the dipole already registers. The small periodic residual  in the 1887 data, treated by Michelson and Morley as an upper bound, reduced by Miller to $U_N \approx 8.8$~km/s \cite{miller1933}, attributed to thermal systematics by later analyses \cite{shankland1955,roberts2006}, and read as signal by others \cite{consoli2004}, converts under this calibration to $U \approx 365$--$375$~km/s, depending on the air's index: bracketing the CMB dipole speed of $369.8$~km/s \cite{planck2020}. Miller's own Mount Wilson amplitude ($U_N \approx 10$~km/s) converts to order $410$~km/s before the apex-projection geometry his reduction requires \cite{miller1933}. We take no position here on whether the residual is signal; the structural point stands either way: the two-law framework converts a century of dispute over these data into one calibration, derived under first-order transport closure, and one discriminating observable.

What the results establish comes in two strengths. At minimum they are exact analog physics: the same laws, to the prefactor, in a system where every part  can be touched, a standing the hydrodynamic analog tradition has already earned \cite{unruh1981,barcelo2011,bush2015}. Beyond that floor, the construction explains: light as the medium's own wave, matter as its closure-bound cavities, the coefficient generated (resonance part by compliant response, share part by closure-forced deformation) rather than postulated. Declaring the correspondence ``mere analogy'' is a metaphysical commitment over and above the mathematics (Section~7); Section~3 documents the cost, paid for a century, of enforcing that commitment as though it had been demonstrated.

\section{Conclusion}

We have re-examined the interpretive history of Fresnel drag with a transport system the nineteenth century did not possess, and the system completes Fresnel's vindication rather than his dissolution. His coefficient contains two laws (Eq.~\ref{eq:template}): the share law he wrote, exact in the compliance language as the carried-compliance fraction, with the Lorentzian composition that reproduces it grounded mechanically by the companion paper; and the resonance law Lorentz added, derived here identically for bubbly liquids and long established for light. The acoustic member has transparent mechanics; what its mechanics reveals is which mechanical channel carries which part. The origin of the resonance part is that waves couple to the most compliant component of structured media and carry the imprint of its resonant response; the origin of the share part is the deformation mechanics of closure-bound cavities: structures whose geometry is fixed by phase closure of the medium's waves.

This equivalence enhances model economy by reducing three apparently independent phenomena to one compliant-structure framework carrying two laws. 
Fresnel drag, Lorentz contraction, and (unexpectedly) atomic polarization and refractive index all emerge from compliant structures deforming under flow or responding to applied gradients.
Moving bubbles exhibit the mechanical repertoire simultaneously: resonance-law entrainment of waves, deformation along the flow direction, and dipole displacement under pressure gradients.

The historical analysis reveals critical junctures:
\begin{enumerate}
\item Fresnel's 1818 choice of density over rigidity shaped the subsequent  interpretive trajectory, before the rigidity alternative was even articulable
\item Kelvin's 1867 vortex-atom model \cite{kelvin1867} supplied structures that could in retrospect have grounded a unification, but it was never applied to the optics of moving bodies
\item Heaviside \cite{heaviside1892}, FitzGerald \cite{fitzgerald1889}, and Lorentz \cite{lorentz1904} derived or proposed aspects of compliant-zone dynamics in 1889--1904, without connecting them to Fresnel's earlier fork
\item Von Laue's 1907 result \cite{vonlaue1907} was received as eliminating mechanics; the verdict is overturned  on two grounds: the resonance law is mechanics under Galilean and Lorentzian kinematics alike, and the deformation law's composition rule is itself derived from closure mechanics \cite{meucci2026cavity}
\end{enumerate}

The two ``compensatory mechanisms'' that critics found suspiciously convenient (Fresnel drag and Lorentz-FitzGerald contraction) are unified consequences when atoms are treated as low-rigidity inclusions. 
The partial drag coefficient $f(\omega)$ and the contraction factor $\sqrt{1-U^2/c^2}$ emerge from the same dynamics.

Following the methodological framework of neoclassical interpretation \cite{meucci2018}, we have shown that mathematical equivalences between formulations need not imply one interpretation is ``correct'' and another ``wrong,'' but rather that multiple perspectives may offer complementary insights. 
The bubble-relativity equivalence, read against the foundations debate on kinematics and dynamics \cite{bell1976,brown2005,janssen2009}, shows that mechanically grounded models remain viable readings of first-order relativistic transport.

The reduction of independent free parameters (one mechanism explaining multiple phenomena) follows the principle of theoretical economy. 
The mathematics is unambiguous about recurrence: the same first-order transport form, and the same Lorentzian closure structure, reappear across the domains. That partial drag, velocity addition, contraction, and polarization are aspects of one principle (compliant inclusions in structured media, deforming under flow) is the mechanical reading this paper constructs. Sufficiency, not uniqueness, is the claim (the mechanics generates the law) and sufficiency is itself the historical point, because the received narrative held that even a mechanical account had been foreclosed in 1907.

\textbf{Derived, measured, and testable.}
What earlier stages of this program stated as a structural prediction (that dispersive corrections to entrainment track the compliance derivative identically in each domain) is now a theorem on the acoustic side and a measurement on the optical side. Appendix~A derives $f_{\mathrm{res}} = (\omega/2)\,C_g'(\omega)/C$ exactly and closure-independently for drifting bubbles; Zeeman's color-resolved interferometry \cite{zeeman1915}, the ring-laser measurement of the dispersive term \cite{bilger1972}, and anomalous-drag experiments in highly dispersive media \cite{boyd2022} supply the optical side, where the same derivative term dominates near resonance. Three tests remain open and are now sharply posed. First, the acoustic law invites direct measurement: impose a controlled bubble drift in liquid at rest and scan frequency through the Minnaert band; the entrainment must follow $(\omega/2)(1/C)\,dC_g/d\omega$ through the Lorentzian derivative structure, with no adjustable constant. The identity is exact within the linear closure model; in real bubbly liquids near resonance, damping, polydispersity, and multiple scattering degrade the idealized propagation model (Commander and Prosperetti report sharply reduced accuracy in resonant regions \cite{commander1989}), so the scan tests the law's derivative structure through the band rather than a parameter-free absolute magnitude. Second, the deformation-law null is itself a prediction: away from resonance the same system must show only a small residue  of order the void fraction: the share-shaped term must be \emph{absent}; no closure in the analyzed Galilean family supplies it, and its presence signals that the compliant response is carried with the flow rather than severed from it; Lorentzian composition reproduces the same value and is itself, by the companion theorem, a product of closure mechanics \cite{meucci2026cavity}. Third (and severable from the first two), Appendix~B turns a century-old dispute into an instrument: a co-located gas-mode/vacuum-mode differential interferometer measures the $(n^2-1)$ residual directly, deciding between first-order transport closure and the second-order completion supplied by exact composition \cite{consoli2004}.

\textbf{Compliance as wave-accessible deformability.}
Compliance in this framework is wave-accessible deformability: the responsiveness of whichever internal degree of freedom stores and returns energy for a given wave type.
For sound, that mode is bubble compression.
For light, electronic polarization.
For a structured-medium vacuum model, substrate deformability.

This naturally explains why, in a bubbly liquid, the compliant component inverts between domains: gas is mechanically deformable; condensed matter is electromagnetically deformable.
The principle holds in each domain; the identity of the compliant subsystem is domain-specific.

Resonance makes the unity explicit.
A gas bubble at Minnaert resonance, an atomic vapor near an electronic transition (as in slow-light experiments achieving group indices exceeding $10^6$ \cite{boyd2022}), and an engineered split-ring resonator all achieve the same result: large effective compliance created by a subwavelength mode whose phase-lagged response modifies the host medium's parameters.
Plasmonic nanoparticles play the electromagnetic role that  bubbles play acoustically: resonant polarizability scaling as inclusion volume in both cases.

The design principle is mechanical: identify the compliant internal mode and tune its resonance to control dispersion and entrainment.
The same Lorentzian derivative structure governs dispersion near resonance whether the mode is acoustic, electric, or magnetic.

For structure, this question is now closed: the companion paper derives the contraction and the period dilation exactly (to all orders in $\beta$) as the unique closure-preserving response of a resonant structure \cite{meucci2026cavity}, completing the direction indicated by the analogue-contraction literature \cite{barcelo2008}. Whether the transport-law equivalence itself extends beyond first order remains open.
This paper and its companion \cite{meucci2026cavity} address complementary halves of the constructive question: how waves are carried by moving structured media, and how wave-organized structures deform when carried. As a study in the history of transport phenomena, the present paper's primary claims are documentary and structural; its physical proposal is calibrated throughout as mechanical sufficiency without uniqueness.
The compliant-inclusion principle, validated independently across acoustic and optical domains, may find application wherever wave propagation through structured media occurs.

\appendix

\section{Two-fluid derivation of the drag coefficient}

\emph{Setup.} Liquid at rest; gas fraction $\alpha$ drifting uniformly at $\mathbf{U}=U\hat{x}$; gas inertia negligible; single pressure field $p$; compliances per volume $C_g = \alpha/K_g$, $C_\ell = (1-\alpha)/K_\ell$, $C = C_g + C_\ell$; liquid density $\rho_\ell$. Perturbations $\propto e^{i(kx-\omega t)}$; unknowns are $p$, the velocity perturbations $u_g, u_\ell$, and the void-fraction perturbation $\alpha'$. Material derivatives: $D_g = \partial_t + U\partial_x$ along the gas, $D_\ell = \partial_t$ along the liquid. The linearized system:
\begin{align}
D_g\alpha' + \alpha\,\partial_x u_g &= -\,C_g(\omega')\,D_g p, \tag{A1}\\
-D_\ell\alpha' + (1-\alpha)\,\partial_x u_\ell &= -\,C_\ell\,D_\ell p, \tag{A2}\\
D_g u_g &= 3\,D_\ell u_\ell, \tag{A3}\\
\partial_t\!\left[(1-\alpha)\rho_\ell u_\ell\right] + D_g\!\left[\tfrac{1}{2}\alpha\rho_\ell(u_g-u_\ell)\right] &= -\,\partial_x p. \tag{A4}
\end{align}
(A1)--(A2) are phase-volume balances with each constitutive response evaluated along its own material path; in (A1) the resonant gas compliance is evaluated at the Doppler frequency the drifting bubbles experience, $\omega' = \omega - kU$, so that $C_g(\omega') \simeq C_g(\omega) - kU\,C_g'(\omega)$ at first order. (A3) is the classical translational response of a massless sphere of added-mass coefficient $\tfrac12$: force balance $\rho_\ell V\,D_\ell u_\ell + \tfrac12\rho_\ell V\,(D_\ell u_\ell - D_g u_g) = 0$ gives acceleration three times the ambient fluid's, each along its own path. (A4) is mixture momentum, the Kelvin impulse of the relative motion transported with the bubbles.

\emph{Extraction.} Nontrivial solutions require the vanishing of the $4\times4$ determinant $D(\omega,k;U)=0$. Writing $\omega/k = v_0 + fU + O(U^2)$ and using the implicit-function theorem, $f = -\,(\partial_U D)/(\partial_\omega D)\,k$ evaluated at $(v_0,\,U{=}0)$. Two sanity checks validate the machinery: (i) with all material derivatives co-moving (the whole mixture drifting), the calculation  returns $f=1$ identically: Galilean full convection; (ii) suppressing the $\alpha'$ advection and the slip reproduces the naive one-channel estimate $f = \tfrac12\,C_g/C$, exposing it as an artifact of the truncation.

\emph{Results.} At $U=0$ the system returns the Wood speed $v_0^2 = 1/[(1-\alpha)\rho_\ell C]$ under the no-slip closure, and its dipole-corrected form $v_0^2 = (1+2\alpha)/(\rho_\ell C)$ under the full closure. At first order in $U$, with frequency-independent compliances, the coefficient is compliance-blind:
\[
f_{0}^{\mathrm{(no\ slip)}} = \frac{\alpha}{2}, \qquad
f_{0}^{\mathrm{(full)}} = \frac{\alpha(11-2\alpha)}{4(1+2\alpha)}\;\approx\;\frac{11\alpha}{4},
\]
with the ratio $C_g/C_\ell$ cancelling identically in both. Restoring the resonant dependence through $C_g(\omega')$ adds, in \emph{both} closures, exactly
\[
f_{\mathrm{res}} = \frac{\omega}{2}\,\frac{C_g'(\omega)}{C}\,,
\]
the closure-sensitive radicals collapsing through the polynomial identity $3a^3+7a^2b+5ab^2+b^3=(a+b)^2(3a+b)$ with $a=\alpha$, $b=1-\alpha$, whose value on the physical simplex $a+b=1$ is simply $1+2\alpha$. The angular dependence generalizes by $U \to U\cos\theta$ throughout. \paragraph{The carried-response model.} The complementary configuration evaluates each component's constitutive response along its own motion: the carried compliance responds in its own frame of reference to the frequency it actually samples, the background responds at rest, and the dispersion relation reads $k^2 = \rho\,[\,C_g(\omega - kU)\,(\omega - kU)^2 + C_\ell\,\omega^2\,]$. The operator choice is the physics: the doubled material derivative on the compliant channel states that the moving component both stores and delivers within its own frame, where the inclusion-drift closures above store in motion but deliver into the mixture at rest. First-order expansion gives, in one line,
\[
f \;=\; \frac{C_g}{C} \;+\; \frac{\omega}{2}\,\frac{C_g'}{C}\,,
\]
the complete first-order coefficient (carried-compliance share plus resonance term) from one constitutive calculation (supplementary script). This is the configuration of the Fizeau experiment under the paper's mapping (compliant inclusions drifting through a background at rest, custody kept through delivery), and it is the calculation Fresnel's own excess-fraction argument performs with density replaced by compliance: the moving excess drifts through the stationary background and carries its share. The inclusion-drift closures above and the carried-response model together bracket the physics: the resonance term is common to every configuration analyzed; the share appears exactly when the response is carried with the flow and disappears when the closure severs response from carrier. At second order the carried-response model and exact velocity composition part company (the optical $O(U^2)$ coefficients differ by a factor of two), the layer Appendix~B addresses. Modern covariant response theory reaches the dispersive coefficient from the opposite direction, by boost-transforming the wavevector- and frequency-dependent matter response \cite{starke2017}; the carried-response relation is its constitutive counterpart, agreeing at first order with no relativistic input.

\paragraph{Closure family and invariance.} The system above is adopted as a linear closure model, with the base slip $U$ externally maintained, rather than derived from a particular averaged two-fluid theory; the family analysis that follows is the answer to that model-dependence. The perturbation throughout is performed at fixed laboratory frequency $\omega$, with the wavenumber $k(U)$ adjusting and the gas compliance sampled at the Doppler-shifted frequency $\omega' = \omega - kU$ of the gas frame; the normalization $k=1$ in the supplementary script is a choice of units, not of physics. The two model closures are members of a wider family: replacing the translational closure by $D_g u_g = q\,D_\ell u_\ell$ and the added-mass impulse by $\mu\,\alpha\rho_\ell D_g(u_g-u_\ell)$, with $(q,\mu)$ arbitrary, and optionally adding the volume-coupled impulse $\tfrac{1}{2}\rho_\ell U\,D_g\alpha'$ motivated by the added-mass analysis of translating variable-volume bubbles \cite{ohl2003}, the resonant coefficient reduces in every case to $f_{\mathrm{res}} = (\omega/2)\,C_g'(\omega)/C$ on the physical simplex $\alpha+(1-\alpha)=1$ (supplementary script). The constant-compliance residue, by contrast, is closure-dependent (the volume-coupled impulse restores $C_g$-dependence to it), but in no member of the family does it take the share form $C_g/C$. A stronger statement holds: solving the share-matching condition across the family yields the unique formal solution $(q,\mu) = \big((\alpha-1)/\alpha,\; 1-\alpha\big)$ in both variants, which requires $q < 0$; no admissible member reproduces the share (supplementary script). The resonance law is invariant over the family; the negative share result is a theorem of the family as a class. The invariance has a structural reading: whenever a drifting channel's resonant constitutive response is sampled at its Doppler-shifted frequency $\omega - kU$, first-order expansion of the phase balance generates the derivative term with prefactor $\omega/2$; the family analysis exhibits that kernel surviving every bookkeeping variation analyzed here.

Near resonance $C_g$ is complex through bubble damping; in the weak-loss regime the real-part derivative carries the Lorentzian structure quoted in Section~2.2, the full complex-wavenumber treatment remaining open.

\section{Gas-mode interferometer sensitivity under the two laws}

This appendix is logically independent of the paper's principal claims; it extends the two laws to one further application.

Let the apparatus and its co-moving gas of refractive index $n$ move at $U$ along $x$ through the substrate. At first order the light's speed in the moving gas is $c/n + fU\cos\theta$ with $f = 1-1/n^2$, and the parallel arm is contracted by $\sqrt{1-U^2/c^2}$ \cite{meucci2026cavity}. The closing speeds along the parallel arm are then $c/n \mp (1-f)U = c/n \mp U/n^2$, and the transverse ray tilts so that its $x$-velocity matches the mirrors':
\begin{align*}
t_\parallel &= \frac{2L\sqrt{1-U^2/c^2}\;(c/n)}{(c/n)^2 - U^2/n^4}\,,\\[2pt]
t_\perp &= \frac{2L}{(c/n)\sqrt{1-\left[(1-f)Un/c\right]^2}}\,.
\end{align*}
Expanding to second order in $U/c$,
\begin{align*}
\Delta t &= t_\parallel - t_\perp \;=\; -\,\frac{L\,U^2\,(n^2-1)}{c^3\,n}\,,\\[2pt]
\Delta N &= \frac{2c\,|\Delta t|}{\lambda} \;=\; \frac{2L\,(n^2-1)}{n\,\lambda}\,\frac{U^2}{c^2}\,.
\end{align*}
Two limits bracket the result. As $n \to 1$ the shift vanishes identically: the deformation law alone nulls the vacuum Michelson--Morley experiment, as Section~8 states. Dropping both drag and contraction returns the classical formula $\Delta N = 2Ln^3U^2/(\lambda c^2)$. A gas-mode instrument therefore retains the classical sensitivity suppressed by $(n^2-1)$: a speed $U_N$ inferred under the classical calibration corresponds to
\[
U \;=\; \frac{U_N}{\sqrt{\,n^2-1\,}}\,,
\]
a factor of approximately $41$ in air. The result is the residual under first-order transport closure: the two laws supply the $O(U)$ propagation speed and the $O(U^2)$ geometry, and no intrinsic $O(U^2)$ transport term is added. Exact velocity composition supplies precisely such a term (its second-order piece restores the covariant null for the co-moving laboratory, gas included), so the second-order gas-mode residual is the discriminating observable between first-order transport closure and exact composition. Three second-order cases are thereby separated by one instrument: the first-order closure above, with no intrinsic second-order transport term (the full residual); the carried-response model of Appendix~A, whose intrinsic second-order term is half the compositional one (half the residual); and exact composition (the null). Whether the completed closure metrology of the companion paper restores the exact null in a co-moving medium at second order is an open derivation, posed in Section~10 alongside the historical record it would adjudicate.

\medskip\medskip
\noindent\textbf{Author contributions:} The author has accepted responsibility for the entire content of this manuscript and approved its submission.

\noindent\textbf{Research ethics:} Not applicable.

\noindent\textbf{Informed consent:} Not applicable.

\noindent\textbf{Use of Large Language Models, AI and Machine Learning Tools:} Large language models were used to assist with drafting, editing, and computational checking under the author's direction; the author reviewed all output and takes full responsibility for the content of the manuscript.

\noindent\textbf{Competing interests:} The author states no conflict of interest.

\noindent\textbf{Research funding:} None declared.

\noindent\textbf{Data availability:} This work is analytical and uses no original experimental data. All numerical values cited are drawn from published sources referenced herein. A commented symbolic-computation script reproducing the derivations of Appendices~A and~B accompanies the manuscript as supplementary material.

\end{document}